# The Digital discrimination of neutron and γ ray using organic scintillation detector based on wavelet transform modulus maximum

YANG Yun(杨云)　　LIU Guo-Fu(刘国福)　　YANG Jun(杨俊) [1)]　　LUO Xiao-Liang(罗晓亮)

College of Mechatronics and Automation, National University of Defense Technology, Changsha 410073, China

**Abstract:** A novel algorithm for the discrimination of neutron and γ-ray with wavelet transform modulus maximum (WTMM) in an organic scintillation has been investigated. Voltage pulses arising from a BC501A organic liquid scintillation detector in a mixed radiation field have been recorded with a fast digital sampling oscilloscope. The performances of most pulse shape discrimination methods in scintillation detection systems using time-domain features of the pulses are affected intensively by noise. However, the WTMM method using frequency-domain features exhibits a strong insensitivity to noise and can be used to discriminate neutron and γ-ray events based on their different asymptotic decay trend between the positive modulus maximum curve and the negative modulus maximum curve in the scale-space plane. This technique has been verified by the corresponding mixed-field data assessed by the time-of-flight (TOF) method and the frequency gradient analysis (FGA) method. It is shown that the characterization of neutron and gamma achieved by the discrimination method based on WTMM is consistent with that afforded by TOF and better than FGA. Moreover, because the WTMM method is itself presented to eliminate the noise, there is no need to make any pretreatment for the pulses.

## 1　Introduction

Neutron detection technology has been widely used in the fields of explosive detection, environmental radiation detection, military and deep space exploration. Since almost all neutron fields coexist with an associated γ-ray component while neutron detectors are also sensitive to γ-ray photons, the discrimination between neutrons and γ rays (n/γ) becomes a key technical problem in the field of neutron detection [1]. Liquid scintillators are one of the most popular radiation detection materials because of their being shaped into the desired size for a specific application and their excellent pulse shape discrimination (PSD) properties and fast timing performance.

A number of existing techniques for PSD can be primarily assorted as time-based and frequency-based. As for time-based techniques, the most popular methods are charge comparison (CC) method [2, 3] and the zero-crossing method [4, 5]. Both were originally implemented in analogue electronics, often in dedicated instrumentation modules or nuclear instrument modules. More recently, they have since been implemented in the digital domain as digital electronic platforms have become available with the requisite speed and cost to make this possible [6, 7, 8]. Currently these methods are the industrial standards to be used to compare with other new proposed discrimination methods, such as the pulse gradient analysis (PGA) [9, 10, 11] method, which is based on the comparison of the relative heights of samples in the trailing edge of the pulse, and curve-fitting method [12]. Since the scintillation process and the photomultiplier tube (PMT) anode signal are often very noisy and time-domain features are naturally highly dependent on the signal amplitude at specific time, these pulse shape discrimination methods can have a great

* Supported by the National Natural Science Foundation of China (A050508/11175254)

1) E-mail: jyang@nudt.edu.cn

dependency on the de-noising algorithm. For example, the results of simulation of PGA show that the efficacy of this method in the presence of significant PMT variation in pulse response is reduced significantly. In the absence of this source of variance to the signals, the figure of merit (FOM) of PGA is theoretically infinite source of variance to the signals.

Based on the above considerations, some researchers have begun to consider PSD methods operating in the frequency domain. In recent years, the frequency-based techniques have emerged as a mature technology, with successful applications across many fields. They are particularly effective pattern recognition tools and can be used to classify neutron and γ-ray events from the measurements performed by organic liquid scintillation detectors. G Liu et al. extracted the frequency-domain features by frequency gradient analysis (FGA) [13, 14] which exhibited a strong insensitivity to the variation in pulse response of the PMT. The FGA method exploits the difference between the zero-frequency component and the first frequency component of Fourier transform of the acquired signal and combines the advantage of insensitivity to noise associated with spectral analysis with that of the real-time implementation of the PGA algorithm. However, for the pulse shapes produced at the output of a liquid scintillator are non-stationary signals, their analysis using conventional Fourier transform could not capture their most relevant features. S.Yousefi et al. adopted a new PSD method based on the wavelet transform [15, 16] to discriminate neutrons and γ rays which presented an efficient tool to analyze non-stationary signals. Experimental results show that compared with PGA algorithm, the wavelet-based method provides an improvement in reducing the overlap between neutron and γ-ray events reflected by an increase in the FOM. Besides, the other methods, such as artificial neural networks [17] and the fuzzy c-means algorithm [18] have been investigated with varying degrees of success.

In this paper, we propose a new PSD method based on wavelet transform modulus maximum (WTMM) which is able to detect neutrons and γ-rays in liquid scintillators. This method exploits the contractive trend between the modulus maximum curve and the modulus minimum curve in the wavelet-domain. In addition, since the WTMM method is itself presented to eliminate the noise, based on the different properties of signals and noise with wavelet transform, there is no need to make any pretreatment for the signals. Moreover, it is verified with characterization of these events arising from the measurement of time of flight (TOF). The TOF technique, which is based on the difference in flight time for neutrons and photons across a known flight path length, has previously been used to compare digital and analogue methods of discrimination and to measure the degree of discrimination quality on a quantitative basis. In this research, it has been used to check the performance of the WTMM. Detailed principle of discrimination with WTMM and the simulation results are given in Section 2; the detailed description of the experimental environment with an associated particle neutron generator and the TOF measurement system are given in Section 3; the experimental results and the comparison with FGA are given in Section 4; the conclusions arising from this research are given in Section 5.

## 2 The principle of the WTMM discrimination system

### 2.1 Empirical characterization of organic liquid scintillator

An empirical formula for the characterization of a liquid scintillation detector for a given volume can provide a generic mathematical description of pulses from that detector which can be used to simulate the response. The six-parameter function [12] that describes scintillator pulse shapes will henceforth be referred to Marrone's model and is given by [19]

$$x(t) = A[\exp(-\frac{t-t_0}{\theta}) - \exp(-\frac{t-t_0}{\lambda_s}) + B\exp(-\frac{t-t_0}{\lambda_l})]. \quad (1)$$

Where A and B are the amplitudes of the short and long components at $t=0$, respectively, $\lambda_s$ is the decay time constant for the short component, while $\lambda_l$ is the long one, $\theta$ is the third decay constant and $t_0$ is the time reference for the start of the signal. Table 1 provides the average values for each of the six parameters for neutron and γ-ray pulse shape using BC501A. Fig.1 shows the normalized amplitude versus time for an average neutron and γ-ray pulse shape according to Marrone's model. The different between neutron and γ ray is evident in the falling portion of each pulse, which is the key part for discriminating.

Table 1. Six-parameter values for Marrone's model for the accurate reproduction of the average neutron and γ-ray pulse shapes (Fig.1) acquired using BC-501A

| Parameter | Radiation type | |
|---|---|---|
| | neutron | γ ray |
| A | 11.48 | 11.03 |
| B | 0.015 | 0.015 |
| $\theta$(ns) | 4.325 | 4.378 |
| $\lambda_s$(ns) | 3.547 | 3.537 |
| $\lambda_l$(ns) | 38.55 | 11.69 |
| $t_0$(ns) | 0.295 | 0.331 |

In this section, the pulse data of neutron and γ ray superimposed on noise with different standard deviation is obtained by Eq. (1) with a sampling rate of 5G samples/s. The WTMM algorithm is used to process these pulse data to discriminate neutron and γ-ray event.

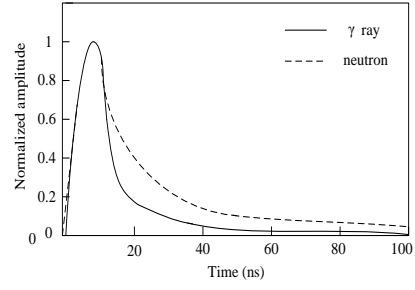

Fig. 1. Normalized pulse of neutron (broken line) and γ ray (continuous line) according to Marrone's model.

## 2.2 The theoretical basis of WTMM

### 2.2.1 Wavelet transform and modulus maximum theory

In the case of nuclear pulse signals, as pulse parameters vary in time and within an ensemble of different pulses, it can be concluded that the process we are dealing with must be considered as nonstationary. In order to analyze this type of pulse, the wavelet transform (WT) uses short time windows at high frequencies and long time windows at low frequencies. Accordingly, the WT provides versions of some parameter estimation at different observation scales. This type of transform converts a function of time, $f(t)$, into a function of two variables: scale and time, $W_f(a,t)$. By evaluating the convolution of the useful and the scaled version of the prototype wavelet, $\varphi_a(t) = \frac{1}{\sqrt{a}}\varphi(\frac{t}{a})$, the transform at different scales may be obtained,

$$W_f(a,t) = f(t) * \varphi_a(t) = \frac{1}{\sqrt{a}}\int_{-\infty}^{+\infty} f(b)\varphi^*(\frac{t-b}{a})db \quad (2)$$

From the point of view of signal response, wavelet transform can be considered as the impulse response of $\varphi_a(t)$. Suppose the wavelet $\varphi_a(t)$ and the signal $f(t)$ to be analyzed are both real functions, and a point $(a_0, t_0)$ in a given scale $a = a_0$, satisfies $\frac{\partial W_f(a,t_0)}{\partial t} = 0$, we call it the local maxima point. Then local modulus maxima at points

can be defined such that

$$|W_f(a_0,t)| \leq |W_f(a_0,t_0)|. \quad (3)$$

where $t$ is either the right or the left neighborhood of $t_0$. Therefore, at a given scale $a_0$, a set of modulus maxima can be obtained from coefficients. Maxima lines are defined as those curves connected by arbitrarily close points in the time-scale plane, along which all points are modulus maxima.

2.2.2 Relationship between modulus maximum and Lipschitz exponent

An important property of the continuous wavelet transform is its ability to react to subtle changes of the signal structure. Of particular importance are the local maxima of $|W_f(a,t)|$ which, as explained above, are the local maxima of the derivative of $f(t)$ smoothed by $\varphi_a(t)$. Mallat and Hwang connected the regularity of a function at a point $t=t_0$ with the decay of the local maxima of the wavelet modulus across scales. Traditionally, the regularity of a function at a point of time or space has been characterized by its Lipschitz exponent

$$|W_f(a,t)| \leq Aa^{\alpha+0.5}. \quad (4)$$

Where $A$ is a constant and $\alpha$ is the Lipschitz exponent. A practical way to calculate the Lipschitz exponent is obviously by rewriting Eq. (4) in the form

$$\log_2(|W_f(a,t)|) \leq \log_2 A + (\alpha+0.5)\log_2 a. \quad (5)$$

Especially, an easy way to obtain Lipschitz exponent is appointed $a = 2^j$, then Eq. (5) is equivalent to

$$\log_2(|W_f(a,t)|) \leq \log_2 A + j(\alpha+0.5). \quad (6)$$

According to the above analysis, there are some peculiar properties between the Lipschitz exponent $\alpha$ and the modulus maxima at a given point $t_0$ of a signal

(1) if $\alpha > -0.5$, with the increase of the scale $j$, the modulus maxima increases.
(2) if $\alpha < -0.5$, with the increase of the scale $j$, the modulus maxima decreases.
(3) if $\alpha = -0.5$, with the increase of the scale $j$, the modulus maxima is invariable.

Eq. (6) and its properties establish the basis of the wavelet method for detecting singularity. The wavelet modulus maxima of singular points, which are effective in the signal, possess translational invariability along different scales and the sign of the values maintain invariable. However, the local modulus maxima of the random noise attenuate rapidly along with the increase of the scales. For singularity detecting, it is generally chosen the derivative of 'Gauss function' as the prototype wavelet to assure modulus maximum curves spreaded to the least scale [20]. In this research, the first derivative of 'Gauss function' wavelet was used.

2.2.3 The discrimination theory of neutron and γ ray based on WTMM

Reference [21] has indicated that the Lipschitz exponent of a Dirac delta function is $\alpha = -1$. From Fig.1, both the normalized neutron waveform and the gamma waveform are similar to the Dirac delta function, so their Lipschitz exponents will be close to -1. However, the decay trend of γ-ray pulse is faster than that of neutron pulse, i.e. the gamma waveform is 'more similar' to the delta function than the neutron waveform, which results in that the Lipschitz exponent of γ-ray pulse is 'more close' to than that of neutron pulse. This slight difference of characteristic information carried by the neutron pulse waveform and the gamma pulse waveform will be used to discriminate them.

To make the above statements clear, the

continuous wavelet transform is computed firstly. Typical neutron and gamma waveforms with their corresponding modulus maximum curves are plotted in Fig.2. Fig.2 (a) and (b) are the normalized pulse of neutron and γ ray obtained from Eq. (1) with a sampling number of 512. Fig.2 (c) and (d) show the continuous waveform function, the bright regions indicate the positive modulus maximum points, and the dark ones are negative modulus maximum points; Fig.2 (e) and (f) are the modulus maximum curves extracted from (c) and (d), respectively.

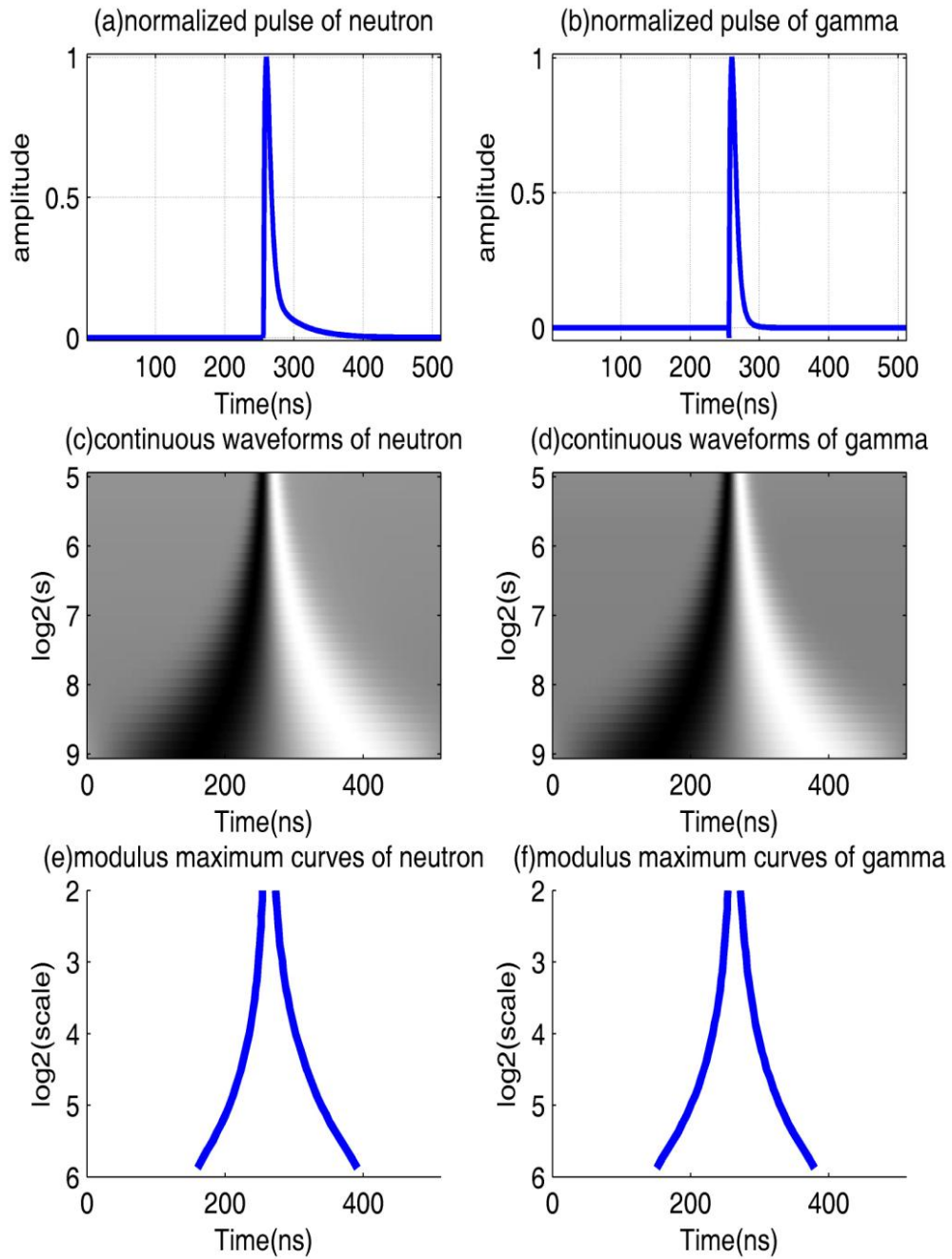

Fig. 2. Typical neutron and gamma waveforms with their corresponding modulus maximum function

The corresponding wavelet coefficients versus scale of the function are shown in Fig.3. The modulus maximum curves of typical neutron and γ ray are inconspicuous in scale-time plane, so we move them to scale-space plane, whose horizontal coordinate is the logarithm of scale, and vertical coordinate is the logarithm of modulus maximum.

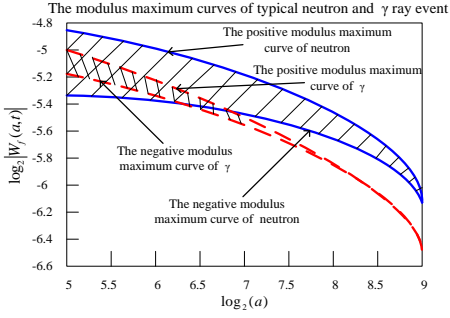

Fig. 3. The modulus maximum curves of typical neutron and gamma. The real lines belong to neutron, and the broken lines belong to γ ray.

We approximately obtain Lipschitz exponents by the curve fitting of a polynomial of degree 1 to each line to calculate the decay ratio in Fig.3 as $\alpha_{\text{n-min}} = -0.686$, $\alpha_{\text{n-max}} = -0.788$, $\alpha_{\gamma\text{-min}} = -0.846$, $\alpha_{\gamma\text{-max}} = -0.891$, which accord with the above statements. It is hard to discriminate γ ray and neutron taking the Lipschitz exponent as a discriminating factor. These calculation results verify the feasibility of the discrimination method based on WTMM. However, the calculation of Lipschitz exponent is not easy, although taking the Lipschitz exponent as the discriminating parameter has a clear meaning,

From Fig.3, we can discover that there are distinct differences between the asymptotic decay trend areas of the γ-ray signal and the neutron signal which can be used as prominent features to discriminate them and easy to compute, so we define a new parameter, i.e. MMT called 'Modulus Maximum Trend', to discriminate neutron and gamma as

$$MMT = \int_{\log_2(scale)} (\log_2|W_f(a,t)|_{\max} - \log_2|W_f(a,t)|_{\min}) d(\log_2(scale)) . \quad (7)$$

The defined MMT is equal to the enwrapped area of the modulus maximum curves corresponding to neutron or γ ray, as shown in the shadows of Fig.3.

### 2.2.4 The discrimination performance of WTMM

In this section, we would like to test the performance of the WTMM method to discriminate neutron and γ ray. Firstly, we mainly investigate how MMT varied with $\lambda_l$ which is the most significant factor of the six-parameter function in Eq. (1) between typical neutron and gamma pulses. Fig.4 shows the change trend between MMT and $\lambda_l$, and their function relation is approximate to

$$MMT = 0.0011\lambda_l^2 + 0.0017\lambda_l + 0.0614 . \quad (8)$$

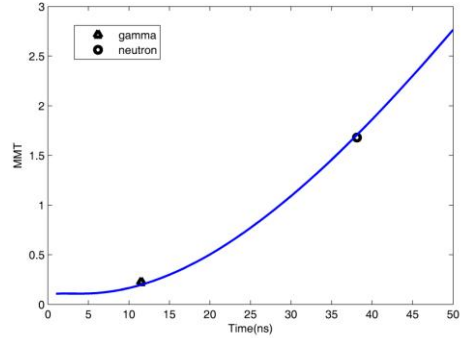

Fig. 4. The change trend between MMT and $\lambda_l$.

In practice, noise always exists in the process of acquiring data, so that denoising becomes a main task in signal processing. With the rapid development of wavelet analysis theory, people pay their attention from spatial domain to wavelet domain. Due to the characteristics of the wavelet such as: multi-scale, multi-resolution analysis, and the improved denoising effect, the denoising method based on wavelet becomes an important research topic in the nuclear field. The sample data of neutron and γ-ray events in the simulation are acquired with various $\lambda_l$ in Eq. (1) with noise $n(t)$. $n(t)$ is a white noise with a normal distribution, with a mean of zero and standard deviation is $\sigma$ which can be varied to change SNR to examine the performance of

WTMM algorithm. Generally the characteristics of signal singularity differ from noise, so do the local maxima of wavelet transform modulus. A denoising algorithm is implemented by their different spread property.

Fig.5 shows MMT varies with $\lambda_l$ under different conditions. From Fig.5 it can be easily observed that with decreasing of SNR less than 11 dB, the error ratios of neutron and γ ray deteriorate sharply. In that conditions, the properties of signal and noise with wavelet transform in scale domain are unconspicuous, the modulus maximum curves of signal are fluctuated by the modulus maximum curves of noise.

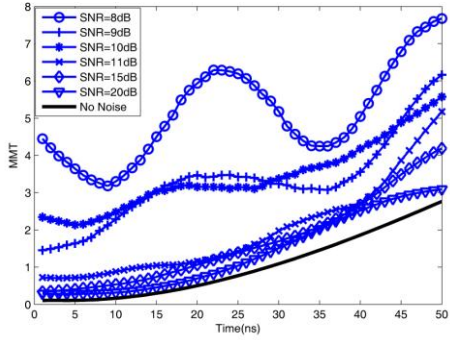

Fig. 5. MMT varies with $\lambda_l$ under different conditions.

## 3  Experimental and TOF results

### 3.1  Experimental setup

The experimental data analyzed in this work were acquired using the TOF measurement system at the Institute of Nuclear Physics and Chemistry, the Chinese Academy of Engineering Physics, Mianyang, China. Fig.6 shows an associated particle neutron generator (APNG), through deuterium-tritium fusion reaction, produces neutrons and alpha particles that are correlated in time and travel in opposite directions to conserve momentum, respectively whose energies are 14.1MeV and 3.5MeV.

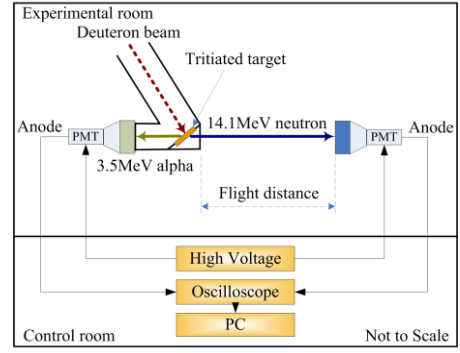

Fig. 6. A schematic of the experimental set-up used for the associated particle neutron generator at the Institute of Nuclear Physics and Chemistry, the Chinese Academy of Engineering Physics, Mianyang, China.

TOF research mainly consists of a plastic scintillation detector and a liquid scintillation detector. Plastic scintillation detector, consisted of a Φ25.4mm×0.1mm cylindrical cell scintillation detector, optically-coupled to an EMI 9807B PMT, which was operated with a negative supply voltage of -1600V DC, detects the arrival of the alpha particle beam pulse and provides a timing reference point for the arrival of each pulse and simultaneity is referred to as the beam-pickup signal. The output signal from the plastic scintillator is connected to Channel 1 of a Tektronic digital phosphor oscilloscope, via approximately 25m of high bandwidth cable. liquid scintillation detector, consisted of a Φ50.8mm×50.8mm cylindrical cell scintillation detector filled with BC501A organic liquid, optically-coupled to another EMI 9807B PMT, which was operated with a negative supply voltage of -1400V DC, is placed at an adjustable distance from the tritiated target detects the corresponding neutrons or γ rays. The output signal from the liquid scintillator is connected to another input of the digital oscilloscope and used to trigger acquisition.

The liquid scintillator pulse and the corresponding beam-pickup pulse data are captured digitally with a sampling rate of 5G Samples/s and 8-bit amplitude resolution. This enabled all detected events, i.e. both γ rays and neutrons, to be sorted in terms of their time-of-arrival relative to the initial beam-pick up.

### 3.2 Experimental results of TOF

TOF spectral analysis is a reliable means of discriminating neutron and γ-ray events, so it is an effective way to verify different discrimination methods. Using the TOF method to discriminate the neutron and γ-ray events, we can verify the WTMM and evaluate the performance of WTMM algorithm. In our research, the distance between the liquid scintillation detector and tritiated target, i.e. the flight path length was set as 1000mm. There are a total of 3679 events that had been recorded and analyzed. The results of TOF method are given in Table 2

Table 2. Experimental results of TOF. There are a total of 3679 events.

| TOF assignment | Neutrons | γ rays | Scatter |
|---|---|---|---|
| Number | 2133 | 1449 | 97 |

According to Table 2, 3582 events are identified as 1449 γ-ray events and 2133 neutron events, but the other 97 events cannot be discriminated and can only be classified as scatter events, which arise as a result of scatter in the environment and within the detector. These scatter events can be either γ-ray events or neutron events. However, the TOF method failed to classify them in terms of their radiation types. The reason for this invalidation of TOF method is that a reference point in time which is usually required in the TOF method is unavailable for scatter events.

The total 3582 pulses induced by these events are applied to the WTMM and FGA method to verify the feasibility of each method respectively. The scatter events are also applied to the WTMM to verify its stability. The discrimination results and discussions of WTMM and FGA method are also shown in the next section.

## 4 Experimental results and discussion

### 4.1 Experimental results of WTMM neutron/gamma discrimination

In the TOF data, there are total of 3769 events that have been classified 2133 neutron events and 1449 γ-ray events. The first 3582 events are discriminated by WTMM; the result is shown in Fig.7. The WTMM method has also been applied to discriminate the scatter events not identified categorically with the TOF method. The results of this analysis, combined with the results of Fig.7, are given in Table 3 and Fig.8. The error of WTMM in discriminating neutrons can be defined as:

$$Error = \frac{|N_{TOF} - N_{WTMM}|}{N_{TOF}}. \quad (9)$$

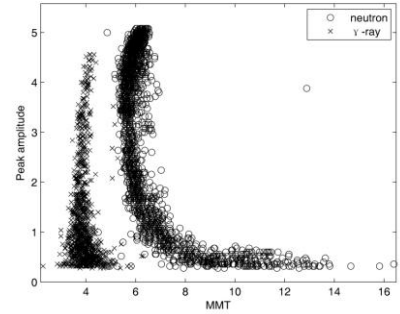

Fig. 7. A total of 3582 events are discriminated by WTMM, the horizontal coordinate is MMT, and the vertical coordinate is the peak amplitude of each pulse.

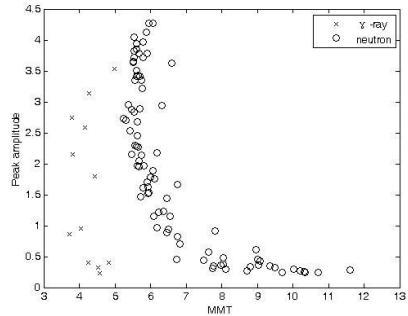

Fig. 8. A total of 97 scatter events are discriminated by WTMM, 12 of which are discriminated as γ rays, the remainings are discriminated as neutrons.

Table 3. The discrimination results of the WTMM method, combined with Table 2

| Method | neutrons | γ rays | scatter |
|---|---|---|---|
| TOF | 2133 | 1449 | 97 |
| WMMT | 2121 | 1433 | 85(n)/12(γ) |
| Error (%) | 0.56 | 1.1 | |

In Section 4.1, we have certified the feasibility of WTMM in the environment of an 8-bit sampling system. There are some events that are erroneously classified by the WTMM method, and they are clearly shown evident in Fig. 7. For example, some γ-ray events are incorrectly tagged by TOF measurement because a neutron event is tagged as such by TOF but is manifest with a γ-ray pulse shape. Because WTMM is essentially a PSD method which only requires the differences of the pulse shape to accomplish n-γ discrimination, it can be used to distinguish the scatter events into two classes, i.e. neutrons and γ rays. In order to further evaluate the overall n-γ separation performance of WTMM, a brief introduction of FGA is presented in section 4.2 and pulses including some induced by scatter events are applied to WTMM compared with FGA to obtain FOM in section 4.3.

### 4.2 Frequency gradient analysis

G. Liu et al. presented a novel n-γ discrimination method called FGA in paper [13], which uses two factors for discrimination after Fourier transform. One is the value at the zero frequency, $|X(0)|$, which is the average value of the signal; the other is the magnitude spectrum of the neutron or γ-ray signal at frequency $f$, $|X(f)|$. Then the discrimination parameter is defined as

$$k(f) = \frac{|X(0)| - |X(f)|}{f}. \quad (10)$$

In the context of the FGA method in this research, we selected the difference between the zero-frequency component and the first frequency component of Fourier transform of the acquired signal, i.e. $|X(0)| - |X(1)|$, as the discrimination parameter to differentiate the neutron event from the γ-ray event.

### 4.3 Figure of merit

Traditionally, the Figure of Merit (FOM) of a discrimination method has relied upon the calculation of a single figure which is used to identify interaction type by simple bounds checking, e.g., if the discrimination figure is over a certain value then it is a particular interaction type. The normalized probability distribution of the discrimination parameter is then obtained and a double Gaussian distribution is fitted to the equation:

$$f(x) = A_1 \exp^{-((x-u_1)^2/2\sigma_1^2)} + A_2 \exp^{-((x-u_2)^2/2\sigma_2^2)}$$

(11)

The FOM is then calculated by

$$\text{FOM} = \frac{S}{\text{FWHM}_\gamma + \text{FWHM}_n}. \quad (12)$$

where S is the separation between the centroids of the neutron peak and the γ-ray peak in the spectrum, and $\text{FWHM}_\gamma$ and $\text{FWHM}_n$ are the Full Width at Half Maximum values of the two peaks of this fit.

Fig.9 shows the distribution for the WTMM discrimination plane together with the double Gaussian fit. As can be seen in Fig.9, there is a good fit between the Gaussian distribution and the probability distribution function, so Eq. (11) and Eq. (12) are used to calculate the FOM of the method.

Table 4. The discrimination results of the WTMM method and the FGA method.

| Method | neutrons | γ rays | FOM |
|---|---|---|---|
| FGA | 1889 | 1055 | 1.135 |
| WTMM | 2121 | 1433 | 1.382 |

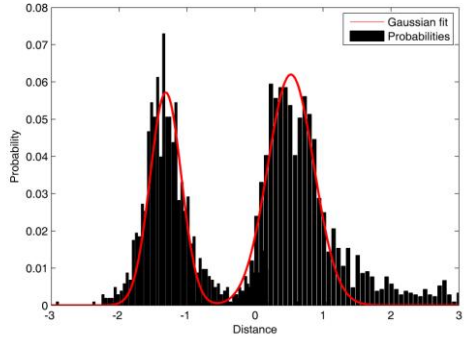

Fig. 9. The distribution of the distance from the mean line between the centroids (of the neutron and gamma plumes) and each points in WTMM discrimination plane.

The discrimination results of the WTMM method and the FGA method are shown in Table 4. The FOMs of FGA and WTMM are calculated as $FOM_{FGA}=1.135$ and $FOM_{WTMM}=1.382$, respectively. For the FOM results, the FOM of WTMM is about 21.8 percent larger than that of FGA, which indicates an improvement of the performance over FGA in discriminating neutrons and γ rays. Moreover, according to the data in Table 4, there are some slight differences in the numbers of each event identified by them and the experimental results obtained from WTMM are slightly closer to those of TOF. For example, the total number of neutron events arising from TOF, FGA, and WTMM are 2133, 1889 and 2121, respectively. There are 244 neutron events mistakenly classified as γ-ray events by FGA, whilst there are only 12 neutron events being mistakenly classified as γ-ray events by WTMM. All these discrimination results indicate directly that the wavelet transform method excelled the Fourier transform method in discriminating neutron and γ ray

## 5 Conclusions

Given the distinct differences of the asymptotic decay trends between the positive modulus maximum curve and the negative modulus maximum curve of neutron and γ-ray signals, which can be used as prominent features to discriminate them, a novel discrimination method based on the wavelet transform called the WTMM method has been proposed. Its discrimination performance has been studied and compared with the TOF method and the FGA method. Theoretical and experiment results show that the WTMM method exhibits a strong insensitivity to the variation in pulse response of the PMT and the electronic noise and therefore takes possession of a better performance than other discrimination methods.

However, the overhead of calculation of this wavelet-based PSD method is heavier than other methods and hence it is not suitable for general real-time embedded systems. However, as the configurable embedded systems continue to develop in capability and speed, the relative merits of wavelet-based methods will be dependent on the elegance with which they are implemented.

*This subject was funded by the National Natural Science Foundation of China. We acknowledge the support of the Institute of Nuclear Physics and Chemistry, the Chinese Academy of Engineering Physics, Mianyang, China. We also appreciate the help and advice of Dr.X.Ma at Department of Engineering, Lancaster University, U.K.*